\documentclass[12pt]{article}
\usepackage{amssymb,amsmath,amscd}
\makeatletter

\@addtoreset{equation}{section}
\def\section{\@startsection {section}{1}{\z@}{-2.25ex plus -1ex minus
 -.2ex}{0.8ex plus .2ex}{\large\bf}}
\def\subsection{\@startsection{subsection}{2}{\z@}{-2.0ex plus%
 -1ex minus -.2ex}{0.4ex plus .2ex}{\bf}}

\def\Ad{\mathrm{Ad}}
\def\ad{\mathrm{ad}}

\newcommand{\inv}[0]{{-1}}

\def\bx{{\mbox{\boldmath $x$}}}

\def\bn{{\mbox{\boldmath $n$}}}

\newcommand{\gothg}{\mathfrak g }

\newcommand{\gothh}{\mathfrak h }

\newcommand{\RR}{\mathbb{R}}
\newcommand{\CC}{\mathbb{C}}

\def\bea{\begin{eqnarray}}
\def\eea{\end{eqnarray}}
\def\bmz{\left(\begin{array}{2,2}}
\def\emz{\end{array}\right)}
\def\bmd{\left(\begin{array}{3,3}}
\def\emd{\end{array}\right)}

\def\bpm{\begin{pmatrix}}
\def\epm{\end{pmatrix}}

\begin{document}
\parskip 6pt
\parindent 0pt
\begin{flushright}
pi-qg-87\\
\end{flushright}

\begin{center}
\baselineskip 24 pt {\Large \bf Quantum double and $\kappa$-Poincar\'e symmetries 
in (2+1)-gravity and Chern-Simons theory}

\baselineskip 16 pt

\vspace{.5cm} {{ C.~Meusburger}\footnote{\tt  cmeusburger@perimeterinstitute.ca}\\
Perimeter Institute for Theoretical Physics\\
31 Caroline Street North,
Waterloo, Ontario N2L 2Y5, Canada\\}

\vspace{0.2cm}

{30 August 2008}

\end{center}

\begin{abstract}
\noindent
We review the role of Drinfeld doubles and $\kappa$-Poincar\'e symmetries in quantised (2+1)-gravity and Chern-Simons theory.  We discuss the conditions under which a given Hopf algebra
symmetry is compatible with a Chern-Simons theory and determine this compatibility explicitly for the Drinfeld doubles and $\kappa$-Poincar\'e symmetries associated with the isometry groups of (2+1)-gravity. In particular, we explain
 that the usual $\kappa$-Poincar\'e symmetries with a timelike deformation are not directly associated with (2+1)-gravity. These $\kappa$-Poincar\'e symmetries are linked to Chern-Simons theory only in the de Sitter case, and the relevant Chern-Simons theory  is physically inequivalent to (2+1)-gravity. 
 \end{abstract}
\centerline{PACS numbers: 04.20.Cv, 02.20.Qs, 02.40.-k}

\section{Introduction and Motivation}

Since their discovery in \cite{LNRT,LNR}, $\kappa$-Poincar\'e symmetries have attracted much interest as possible symmetries of a quantum theory of gravity in three and higher dimensions. The idea is that the usual Poincar\'e symmetry of Minkowski space or, more generally, the isometry groups of de Sitter and anti de Sitter space,  are deformed into Hopf algebras. These Hopf algebras, which have been shown to have the structure of a bicrossproduct \cite{MR},  depend on a parameter $\kappa$ with the dimension of a mass and hence give rise to an invariant mass scale in the theory. 

Although motivated by phenomenological considerations, the discussion of $\kappa$-Poincar\'e symmetries in four dimensions remains largely heuristic due to the lack of a complete quantum theory of gravity. Moreover, their physical  interpretation is complicated by the absence of a derivation from an action functional and a  lack of $\kappa$-Poincar\'e invariant models with non-trivial interactions.

In (2+1) dimensions, the situation simplifies considerably, both with respect to the  quantisation of gravity and to the role of Hopf algebras as quantum symmetries. This is due to the fact that  (2+1)-dimensional gravity can be formulated as a Chern-Simons gauge theory in whose quantisation Hopf algebra structures are known to arise.  The (2+1)-dimensional case has thus served as a toy model for the motivation and interpretation of $\kappa$-Poincar\'e symmetries in higher dimensions.

However, the Hopf algebras which are well-established as a quantum symmetries of (2+1)-gravity 
are not  $\kappa$-Poincar\'e symmetries,
 but a different set of deformations
 of the isometry groups of (2+1)-gravity, the Drinfeld doubles 
$D(U(\mathfrak{su}(1,1)))$, $D(U_q(\mathfrak{su}(1,1)))$, $q\in\RR$, and $D(U_q(\mathfrak{su}(1,1)))$, $q\in U(1)$, for,  respectively, vanishing, positive and negative cosmological constant.
While these Drinfeld doubles are known to arise in the quantisation of the theory \cite{BNR, we2}, the status of $\kappa$-Poincar\'e symmetries is less clear. Previous work investigating their role in (2+1)-gravity focussed solely on the algebra structure while neglecting the coalgebra, which in itself is not sufficient for establishing the presence of a Hopf algebra symmetry\footnote{It can be shown that both the Drinfeld double $D(U(\mathfrak{su}(1,1)))$ and the $\kappa$-Poincar\'e algebra are isomorphic to the (2+1)-dimensional Poincar\'e algebra as {\em algebras}. Hence, the algebra structure alone does not allow one to distinguish these Hopf algebras from each other and from the (2+1)-dimensional Poincar\'e algebra. }.  The situation is complicated further by the fact that both types of deformations share certain physical features such as non-commutative position coordinates and group valued momentum variables. Mathematically, both are associated with the factorisation of the isometry groups of (2+1)-gravity into the 
(2+1)-dimensional Lorentz group $SU(1,1)$ and the group $AN(2)$. 

The purpose of this article is to review the status of Drinfeld doubles and $\kappa$-Poincar\'e symmetries as symmetries of quantised (2+1)-gravity. We explain how the presence or absence 
of these Hopf algebras in quantised (2+1)-gravity can be determined
 by considering their classical limit in the Chern-Simons formulation of (2+1)-gravity \cite{wekappa}.  It is well known that the Hopf algebras arising in the quantisation of Chern-Simons theory are the quantum counterparts of certain Poisson-Lie structures which describe the Poisson structure on the phase space of the classical theory.  By considering the classical limit, one is thus able to reduce the question about the  presence of these Hopf algebra symmetries to the analogous question for the associated classical and infinitesimal structures, Poisson-Lie groups and  Lie bialgebras. 
 Using the results of \cite{wekappa,we6}, we address this question for the Poisson-Lie structures and Lie bialgebras associated with the Drinfeld doubles and a generalised version of the $\kappa$-Poincar\'e symmetries. 

The paper is structured as follows: In Sect.~\ref{halgsymms} we summarise the relevant facts on Hopf algebra symmetries, Poisson-Lie groups and Lie bialgebras.  Sect.~\ref{CSsect} gives a brief overview of Chern-Simons theory with a focus on the Chern-Simons formulation of (2+1)-gravity. In
Sect.~\ref{plcs}, we discuss the role of Poisson-Lie structures in the description of the phase space of the  theory and explain how one can determine if a given Poisson-Lie structure and the associated Hopf algebra are compatible with a given Chern-Simons theory. 
In Sect.~\ref{kapsect}, we apply this procedure to the Poisson-Lie structures associated with the
Drinfeld doubles and a generalised version of $\kappa$-Poincar\'e symmetries, whose role in (2+1)-dimensional gravity has been investigated in, respectively, \cite{we6} and \cite{wekappa}. 
 Sect.~\ref{conc} contains our outlook and conclusions.

\section{$\kappa$-Poincar\'e and quantum double symmetries as Hopf algebra symmetries}

\label{halgsymms}

Both $\kappa$-Poincar\'e symmetries and Drinfeld doubles are examples of Hopf algebra symmetries which generalise the usual notion of group symmetries in physics.  A Hopf algebra is an associative algebra $\mathfrak G$ with additional structures, the coproduct $\Delta: \mathfrak G \rightarrow \mathfrak G\otimes\mathfrak G$, the counit $\epsilon: \mathfrak G\rightarrow \CC$ and the antipode $S:\mathfrak G\rightarrow \mathfrak G$ which satisfy certain consistency conditions and compatibility conditions  with the algebra structure.  The role of the coproduct is that it allows one to define representations of $\mathfrak G$ on the tensor product of representation spaces of $\mathfrak G$,  which can be thought of as multi-particle states. The antipode gives rise to representations of $\mathfrak G$ on the duals of its representation spaces, which are usually interpreted as anti-particles. 
Group symmetries in quantum systems  present  a trivial example of Hopf algebra symmetries, in which the Hopf algebra is the universal enveloping algebra of the associated Lie algebra $\mathfrak G=U(\gothg)$.  Coproduct and antipode take the form
$\Delta(\bx)=1\otimes \bx=\bx\otimes 1$, $S(\bx)=-\bx$ $\forall \bx\in\gothg$, and one recovers the familiar formulas for the symmetries associated with multi-particle states and antiparticle systems.

Hopf algebras can be viewed as the quantum counterparts of corresponding classical structures, namely Poisson-Lie groups. A Poisson-Lie group is a Lie group $G$ equipped with a Poisson structure such that the multiplication $G\times G\rightarrow G$
is a Poisson map. For the trivial case of universal enveloping algebras discussed above, this Poisson structure is the trivial one. 
Poisson-Lie group structures  are in turn defined by associated infinitesimal structures,  Lie bialgebras. A Lie bialgebra $\mathfrak g$ is a Lie algebra with an additional structure, the cocommutator $\delta:\mathfrak g\rightarrow \mathfrak g\otimes\mathfrak g$ that satisfies certain compatibility condition with the Lie bracket. It can be viewed as the infinitesimal counterpart or first- 
order term of the coproduct $\Delta: \mathfrak G\rightarrow \mathfrak G\otimes\mathfrak G$, while the Lie bracket corresponds to the first order term of the commutator of $\mathfrak G$. 

For quasi-triangular  Hopf algebras such as the Drinfeld doubles and $\kappa$-Poincar\'e symmetries, the corresponding Lie bialgebra structures are in turn defined by an additional structure, the classical $r$-matrix for the Lie algebra $\mathfrak g$. This is an element
$r=r^{\alpha\beta}X_\alpha\otimes X_\beta \in \mathfrak g \otimes\mathfrak g$
which satisfies the classical Yang Baxter equation (CYBE)
\begin{align}
\label{CYBE}
[[r,r]]=[r_{12}, r_{13}]+[r_{12}, r_{23}]+[r_{13}, r_{23}]=0
\end{align}
where  $r_{12}=r^{\alpha\beta} X_\alpha\otimes X_\beta\otimes 1$, $r_{13}=r^{\alpha\beta} X_\alpha\otimes 1\otimes  X_\beta$, $r_{23}=r^{\alpha\beta} 1\otimes X_\alpha\otimes X_\beta$ and $X_\alpha$, $\alpha=1,\ldots, \text{dim}\,\mathfrak g$, is a basis of $\mathfrak g$.

The relationship between Hopf algebras, Poisson-Lie groups, Lie bialgebras and classical $r$-matrices is illustrated in Table 1:  Via the identity
\begin{align}
\label{cocomm}
\delta(X_\gamma)=(1\otimes \ad_{X_\gamma}+\ad_{X_\gamma}\otimes 1)(r)=r^{\alpha\beta}\left( X_\alpha\otimes[X_\gamma, X_\beta] + [X_\gamma,X_\alpha]\otimes X_\beta\right)
\end{align}
 the classical $r$-matrix  defines the cocommutator of $\mathfrak g$ and hence its Lie bialgebra structure. Exponentiation then yields the associated Poisson-Lie group and quantisation the corresponding Hopf algebra.

An important notion is the concept of Hopf algebra duality. Both the $\kappa$-Poincar\'e and the Drinfeld double symmetries considered in this paper correspond to a pair of Hopf algebras in duality.
For the $\kappa$-Poincar\'e symmetries these are the $\kappa$-Poincar\'e algebra and its dual, the $\kappa$-Poincar\'e group \cite{KM}. In the case of Drinfeld double symmetries, one has the Drinfeld double and its dual. 
Two Hopf algebras are in duality if they are dual as vector spaces, their antipodes are in duality   and the comultiplication and counit  of one of them are dual to, respectively,  the multiplication and unit of the other. In the quasi-triangular case, the Poisson-Lie structures corresponding to such a pair of dual Hopf algebras are a pair of Poisson-Lie groups in duality, the Sklyanin Poisson structure and its dual, the dual Poisson structure. The infinitesimal counterpart of such a pair of  Poisson-Lie groups in duality is a pair of dual Lie bialgebras, i.~e.~ two Lie algebras which are dual as vector spaces and such that the Lie bracket of one is  dual to the cocommutator of the other and vice versa. 
These notions of duality are illustrated in Table 1.

For the purpose of this paper, it is important to note that the classical $r$-matrix defines all of the structures introduced above. This implies that a quasi-triangular Hopf algebra is essentially determined by the corresponding classical $r$-matrix and allows one to reduce the question about the role of Hopf algebras as symmetries of quantised (2+1)-gravity to a question about the compatibility  of the associated classical $r$-matrices with the action functional defining  the theory.

\vbox{
\label{halgtable}
\baselineskip 38 pt
\begin{center}

\footnotesize

\begin{tabular}{l|lll}
{\bf Hopf algebras} &{\bf Hopf algebra $\mathfrak G$} & $\leftarrow${\bf Hopf algebra duality}$\rightarrow$ & {\bf dual Hopf algebra $\mathfrak G^*$}\\
\\[-2ex]
& $\kappa$-Poincar\'e algebra & & $\kappa$-Poincar\'e group\\
& Drinfeld doubles & & dual of Drinfeld doubles\\
\\[-2ex]
\hline
\\
\multicolumn{4}{c} {$\uparrow$ quantisation: $\quad\{\,,\,\}\rightarrow i\hbar [\,,\,]$, $\delta\rightarrow \Delta$}\\
\\
\hline
\\[-2ex]
{\bf classical} & {\bf dual Poisson } & $\leftarrow${\bf Poisson-Lie group duality}$\rightarrow$ & {\bf Sklyanin bracket } \\
{\bf counterparts:}& {\bf structure on $G$} & &{\bf on $G$}\\
{\bf Poisson-Lie } \\
{\bf groups}\\
\\[-2ex]
\hline
\\
\multicolumn{4}{c}{$\uparrow$ exponentiation}\\
\\
\hline
\\[-2ex]
{\bf infinitesimal } & {\bf Lie bialgebra $\mathfrak g$} & $\leftarrow${\bf duality of Lie bialgebras}$\rightarrow$ & {\bf dual Lie bialgebra $\gothg^*$} \\
{\bf counterparts:  }\\
{\bf Lie bialgebras} & $[X_\alpha, X_\beta]=f_{\alpha\beta}^{\;\;\gamma} \, X_\gamma$ & &$\delta_*(X^\alpha_*)\!=\!f_{\beta\gamma}^{\;\;\alpha} X^\beta_*\otimes X^\gamma_*$\\
& $\delta(X_\alpha)\!=\!g_\alpha^{\;\beta\gamma} X_\beta\otimes X_\gamma$ & & $[X_*^\alpha, X_*^\beta]_*=g_{\gamma}^{\;\alpha\beta} X^\gamma_*$\\
\\[-2ex]
\hline
\\
\multicolumn{4}{c}{$\uparrow$ cocommutator $\quad\delta(X_\alpha)=(1\otimes \ad_{X_\alpha}+\ad_{X_\alpha}\otimes 1)(r)$}\\
\\
\hline
\\[-2ex]
{\bf classical $r$-matrix} & \multicolumn{3}{l} {$r=r^{\alpha\beta}X_\alpha\otimes X_\beta\in \mathfrak g \otimes\mathfrak g$}\\
\end{tabular}

\vspace{0.4cm}

\normalsize
Table 1:  Hopf algebras, Poisson-Lie groups and Lie bialgebras \end{center}
}

\section{The Chern-Simons formulation of (2+1)-gravity}

\label{CSsect}

As shown in \cite{AT,Witten1}, (2+1)-dimensional gravity can be formulated as a Chern-Simons gauge theory. The two ingredients needed in the definition of a  Chern-Simons gauge theory are a Lie group $G$ and a non-degenerate symmetric bilinear form on $\mathfrak g=\text{Lie}\;G$ that is invariant under the adjoint action $\Ad$ of $G$ on $\gothg$. 
Given a gauge group $G$ and an $\Ad$-invariant symmetric bilinear form $\langle\,,\,\rangle$ on $\gothg$, the corresponding Chern-Simons theory on a three-manifold $M$ is then given by Chern-Simons action
\begin{align}
\label{CSact}
S[A]=\int_M \langle dA\wedge A+\tfrac{2}{3} A\wedge A\wedge A\rangle,
\end{align}
where $A$ is the Chern-Simons gauge field which is locally a one-form on $M$ with values in the Lie algebra $\gothg$ and transforms under gauge transformations $\gamma: M\rightarrow G$ as $A\mapsto \gamma A \gamma^\inv+\gamma d\gamma^\inv$.

In the Chern-Simons formulation of (2+1)-gravity, the gauge groups are the isometry groups of the generic solutions of the theory, three-dimensional Minkowski space, de Sitter space and anti de Sitter space.  Adapting the conventions from \cite{Witten1} and  denoting by  $\Lambda$ minus the cosmological constant, one finds that these isometry groups are the
three-dimensional Poincar\'e group $ISO(2,1)=SO(2,1)\ltimes\RR^3$ for $\Lambda=0$, the group $SO(2,1)\times SO(2,1)$ for $\Lambda>0$ and the group  $PSL(2,\CC)$ for $\Lambda<0$ as shown in Table 2.

\vbox{
\label{cctable}
\baselineskip 38 pt
\begin{center}
\begin{tabular}{|c|c|c|}
\hline
$\Lambda$   & spacetime $M_\Lambda$ & isometry group $H_\Lambda$\\
\hline
$ =0$  & $ \mathbb M^3$ & $SO(2,1)\ltimes \RR^3$ \\
\hline
$> 0$ &
$\text{AdS}_3$&
$SO(2,1)\times SO(2,1)$ \\
\hline
$ < 0$ & $ \text{dS}_3 $&
$PSL(2,\mathbb C) $\\
\hline
\end{tabular}
\vspace{0.1cm}

Table 2: Generic spacetimes and isometry groups in (2+1)-gravity
\end{center}
}

The associated Lie algebras  $\gothh_\Lambda=\text{Lie}\; H_\Lambda$ take the form
\begin{align}
\label{liebra} &[J_a,J_b]=\epsilon_{abc} J^c & 
&[J_a,P_b]=\epsilon_{abc} P^c &
&[P_a,P_b]=\Lambda\epsilon_{abc}J^c,
\end{align}
where  $J_a$, $a=0,1,2$, are the generators of the Lorentz transformations, $P_a$, $a=0,1,2$, the generators of the translations, 
$\epsilon^{abc}$ is the totally antisymmetric tensor satisfying  $\epsilon_{012}=\epsilon^{012}=1$ and indices are raised and lowered with the three-dimensional Minkowski metric $\eta=\text{diag}(1,-1,-1)$. 

The other ingredient in the Chern-Simons formulation of (2+1)-gravity is an $\Ad$-invariant non-degenerate symmetric bilinear form on $\gothh_\Lambda$. While the choice of this
 form is unique for  semi-simple gauge groups, the space of $\Ad$-invariant symmetric bilinear forms is two-dimensional for the Lie algebras  $\gothh_\Lambda$. As shown in \cite{Witten1}, for all signs of $\Lambda$ a basis is given by the forms
\begin{align}
\label{pair} &t( J_a,J_b)=0, & &t(P_a,P_b)=0, &
&t( J_a,P_b)=\eta_{ab},\\
\label{othform} &s(J_a,J_b)=\eta_{ab}, & &s(J_a,P_b)=0, &
&s(P_a,P_b)=\Lambda\eta_{ab}.
\end{align}
A general $\Ad$-invariant symmetric bilinear form $\tau$ on $\gothh_\Lambda$ is a linear combination
$\tau=\alpha t +\beta s$, $\alpha,\beta\in\RR$. The requirement of non-degeneracy then takes the form  $\alpha^2-\Lambda\beta^2\neq 0$. Note in particular that, while the form $t$ is non-degenerate for all signs of $\Lambda$, the form $s$ becomes degenerate for $\Lambda=0$.

As shown in \cite{AT, Witten1}, the Chern-Simons formulation of (2+1)-gravity  is given by the action \eqref{CSact} with $G=H_\Lambda$, $\langle\,,\,\rangle=t$ and is obtained by combining the triad  $e^a=e^a_\mu dx^\mu$ and the spin connection $\omega^a=\omega^a_\mu dx^\mu$ in the first order formulation of (2+1)-gravity into a Chern-Simons gauge field
\begin{align}
\label{gfield}
A(x)=\omega^a(x)J_a+e^a(x)P_a.
\end{align}
Using  \eqref{CSact},  \eqref{gfield} and  \eqref{pair}, one can show that the Chern-Simons action \eqref{CSact} agrees with the Einstein-Hilbert action for (2+1)-gravity in the first order formulation and that the metric on $M$ is given by $g_{\nu\mu}=\eta_{ab}e^a_\mu e^b_\nu$.

The equations of motion derived from the action \eqref{CSact} are a flatness condition on the gauge field $A$, which  combines the requirements of vanishing torsion and constant curvature
\begin{align}
\label{equofmot}
F(A)=dA+\tfrac{1}{2}[A\wedge A]=0\quad\Leftrightarrow\quad &T^a=de^a+\epsilon^{abc}\omega_b\wedge e_c=0\\
& R^a=d\omega^a+\tfrac{1}{2}\epsilon^{abc}\omega_b\wedge\omega_c=-\tfrac{\Lambda}{2}\epsilon^{abc}e_b\wedge e_c\nonumber.
\end{align}
It can be shown that the equations of motion are independent of the choice of the $\Ad$-invariant symmetric bilinear form on $\gothh_\Lambda$.  However, it is important to stress that different choices of this form nevertheless define physically non-equivalent theories as they affect its Poisson structure and physical interpretation. 
As we will see in the following that it is precisely the choice of this form  that  determines if the resulting quantum theory will exhibit  $\kappa$-Poincar\'e or  Drinfeld double symmetries, understanding how different choices of this form affect the classical theory is crucial for the central aim of this paper. 

To clarify the impact of the $\Ad$-invariant, symmetric bilinear form, it is instructive to focus on the extreme cases $\tau=\alpha t$, $\beta=0$  (the form corresponding to (2+1)-gravity)  and $\tau=\beta s$, $\alpha=0$.   Considering a manifold of topology $M\approx\RR\times S$ and performing a (2+1)-decomposition of the gauge field, one finds that the Poisson structure
for the choice $\tau=\alpha t$ is given by 
\begin{align}
\label{symp1}
\{e_i^a(x), \omega_j^b(y)\}=\frac{\eta^{ab}}{2\alpha}\epsilon_{ij}\delta_S(x-y)\qquad \{e_i^a(x),e_j^b(y)\}=\{\omega_i^a(x),\omega_j^b(y)\}=0
\end{align}
while the other $\Ad$-invariant symmetric bilinear form $\tau=\beta s$ yields
\begin{align}
\label{symp2}
\{e_i^a(x), e_j^b(y)\}\!=\!\frac{\eta^{ab}}{2\Lambda\beta}\epsilon_{ij}\delta_S(x-y),\;\{\omega_i^a(x), \omega_j^b(y)\}\!=\!\frac{\eta^{ab}}{2\beta}\epsilon_{ij}\delta_S(x-y),\;  \{e_i^a(x),\omega_j^b(y)\}\!=\!0.
\end{align}
The impact on the physical interpretation of the theory is directly apparent when the theory includes point particles minimally coupled to the Chern-Simons gauge field. As shown in \cite{wekappa}, 
 exchanging the Poisson brackets \eqref{symp1} and \eqref{symp2} amounts to exchanging the internal degrees of freedom of the particles, their mass and internal angular momentum or spin. A similar effect arises when one considers spacetimes with a boundary corresponding to an asymptotic observer, where switching the \eqref{symp1}, \eqref{symp2} amounts to exchanging the asymptotic degrees of freedom, the total energy and total angular momentum of the universe. 
 
 However, the effect is also present for purely topological spacetimes without matter or boundaries \cite{wekappa}. In this case, it can be shown that each closed, non-selfintersecting curve on the spatial surface $S$ is equipped with two canonical observables, which can be viewed as a ``mass'' or momentum and a ``spin'' or angular momentum  and generalise the associated quantities for particles. Via the Poisson bracket, these observables generate earthquakes (cutting the spatial surface along the curve and rotating the edges of the cut against each other) and grafting (cutting the spatial surface along the curve and inserting a cylinder), which can be viewed as, respectively, a rotation and a translation associated to the curve.  Exchanging  the Poisson brackets \eqref{symp1}, \eqref{symp2} amounts to switching the two canonical observables with respect to this geometry changing transformations  and thus affects their physical interpretation.

\section{Hopf algebra symmetries and Poisson-Lie structures in Chern-Simons theory}

\label{plcs}

The advantage of the Chern-Simons formulation of (2+1)-gravity is that it gives rise to a simple and efficient description of the phase space and Poisson structure of the theory. The phase space of Chern-Simons theory with gauge group $G$ on a manifold $M\approx \RR\times S_{g,n}$, where $S_{g,n}$ is an oriented two-surface of genus $g$ and with $n$ punctures representing point particles,  is the quotient of the space of holonomies along a set of generators of the fundamental group $\pi_1(S_{g,n})$ modulo simultaneous conjugation 
\begin{align}
\mathcal M_{g,n}(G)=\text{Hom}(\pi_1(S_{g,n}), G)/G.
\end{align}
This parametrisation of the phase space establishes a direct relation between its Poisson structure and the theory of Poisson-Lie groups. As shown in \cite{FR}, the 
 the Poisson structure on the phase space $\mathcal M_{g,n}(G)$ can be described in terms of an auxiliary Poisson structure on the manifold $G^{n+2g}$ defined in terms of a classical $r$-matrix for $G$. Moreover, it has been demonstrated in \cite{AMII} that the contributions of each puncture 
  and handle of $S_{g,n}$ to this Poisson structure can be decoupled and then correspond, respectively, to a copy of the dual Poisson structure and of the Heisenberg double Poisson structure defined by this classical $r$-matrix.

The classical $r$-matrices arising in this description of the phase space  are general solutions of the CYBE \eqref{CYBE} which must satisfy a certain compatibility condition relating them to the Chern-Simons action. This condition states that their symmetric component $r_s=\tfrac{1}{2}(r^{\alpha\beta}+r^{\beta\alpha})X_\alpha\otimes X_\beta$ is dual to the $\Ad$-invariant symmetric bilinear form in the Chern-Simons action which defines the Poisson structure of the theory.

The Hopf algebra  symmetries of the quantised Chern-Simons theory  then arise as the quantum counterparts of the Poisson-Lie  and Lie bialgebra structures defined by this classical $r$-matrix. Hence, assuming the existence of a well-defined classical limit,  one can use the link between the Poisson structure on phase space and classical $r$-matrices to determine explicitly if a given quasi-triangular Hopf algebra arises as a quantum symmetry of a given Chern-Simons theory. For this, one takes the following steps:\vspace{-.3cm}
\begin{enumerate}
\item Determine the associated Poisson-Lie structures and Lie bialgebras via the classical limit.\vspace{-.3cm}
\item  Determine the antisymmetric component $r_a=\tfrac{1}{2}(r^{\alpha\beta}-r^{\beta\alpha})X_\alpha\otimes X_\beta$ of the classical $r$-matrix defining these structures via \eqref{cocomm}.\vspace{-.3cm}
\item  Consider the symmetric element  $r_s\in \gothg\otimes\gothg$ corresponding to the $\Ad$-invariant symmetric form in the Chern-Simons action.\vspace{-.3cm}
\item Determine if the sum $r_s+r_a$ satisfies the CYBE \eqref{CYBE}.
\end{enumerate}

\section{Drinfeld doubles and $\kappa$-Poincar\'e symmetries in (2+1)-gravity and Chern-Simons theory}

\label{kapsect}

We now apply this procedure  to determine the role of Drinfeld double and $\kappa$-Poincar\'e symmetries as quantum symmetries of  (2+1)-gravity and, more generally, Chern-Simons theory with gauge group $H_\Lambda$. A detailed investigation of the Lie bialgebra  and Poisson-Lie group structures underlying the  $\kappa$-Poincar\'e symmetries  and Drinfeld doubles is given in  \cite{BBH,BHDS,we6}, and the compatibility of $\kappa$-Poincar\'e symmetries with Chern-Simons theories with gauge group $H_\Lambda$ is investigated in \cite{wekappa}. 

Using the results of \cite{BBH,BHDS,we6}, one can  determine the antisymmetric component of the classical $r$-matrix which defines these Lie bialgebra structures  via \eqref{cocomm}. As shown in \cite{we6}, for all values of the cosmological constant, the antisymmetric component of the classical $r$-matrix defining the Drinfeld doubles  is of the form
\begin{align}
\label{radouble}
r_a=\tfrac{1}{2}(P_a\otimes J^a-J_a\otimes P_a)+n^a\epsilon_{abc} J^b\otimes J^c\qquad \bn\in\RR^3.
\end{align}
For the case of  $\kappa$-Poincar\'e symmetries we have from \cite{BBH,BHDS}
\begin{align}
\label{rakappa}
r_a=n^a\epsilon_{abc}(J^b\otimes P^c-P^c\otimes J^b)\qquad \bn\in\RR^3
\end{align}
with the standard $\kappa$-Poincar\'e symmetries corresponding to vanishing cosmological constant and a timelike deformation with
$\bn=(\tfrac{1}{\kappa}, 0,0)$. 
The symmetric element of $\gothh_\Lambda\otimes\gothh_\Lambda$ dual to a  general $\Ad$-invariant bilinear form $\tau=\alpha t+\beta s$ on the Lie algebra $\gothh_\Lambda$
is \cite{wekappa}
\begin{align}
\label{rsgen}
r_s=\frac{\alpha}{\alpha^2-\Lambda\beta^2}(P_a\otimes J^a+J_a\otimes P^a)-\frac{\beta}{\alpha^2-\Lambda\beta^2}(\Lambda\, J_a\otimes J^a+P_a\otimes P^a).
\end{align}
 Recalling the discussion after \eqref{othform}, we note that the normalisation factor  $1/\alpha^2-\Lambda\beta^2$ arising from the duality condition
 diverges precisely when the  form $\tau$ becomes degenerate.
 The requirement that the sum $r_a+r_s$  yields a solution of the  CYBE \eqref{CYBE} then results in conditions on the coefficients $\alpha,\beta$ in the $\Ad$-invariant symmetric bilinear form and on the vector $\bn$ in \eqref{radouble}, \eqref{rakappa}. For the Drinfeld doubles  \eqref{radouble} these conditions imply that the solutions of the CYBE must satisfy
$\beta=0$, $\bn^2=-{\Lambda}/{\alpha^2}$. The classical $r$-matrices are then given by
\begin{align}
\label{rmatdouble}
r=r_a+r_s=\frac{1}{\alpha}P_a\otimes J^a+n^a\epsilon_{abc}J^b\otimes J^c \qquad \bn^2=-\frac \Lambda {\alpha^2}.
\end{align}
The Drinfeld doubles therefore correspond to the Chern-Simons formulation of (2+1)-gravity determined by the $\Ad$-invariant symmetric bilinear form \eqref{pair}. The vector $\bn$ is timelike, spacelike and lightlike or zero, respectively, for $\Lambda<0$ (de Sitter case), $\Lambda>0$ (anti de Sitter case) and $\Lambda=0$ (Minkowski case).  This implies that the Drinfeld doubles are compatible  with quantised (2+1)-gravity for all signs of the cosmological constant.

It is shown in \cite{wekappa} that the requirement that $r_s+r_a$ satisfies the CYBE \eqref{CYBE} yields  two sets of solutions for the generalised  $\kappa$-Poincar\'e symmetries \eqref{rakappa} . The first is characterised by the conditions $\beta=0$, $\bn^2=-{1}/{\alpha^2}$ and the classical $r$-matrices
\begin{align}
\label{rmatkappa1}
r=r_a+r_s=\frac{1}{\alpha}(P_a\otimes J^a+J_a\otimes P^a)+n^a\epsilon_{abc}(J^b\otimes P^c-P^c\otimes J^b)\qquad \bn^2=-\frac{1}{\alpha^2}.
\end{align}
It corresponds again to the Chern-Simons formulation of (2+1)-gravity with $\Ad$-invariant symmetric bilinear form \eqref{pair} and involves a spacelike vector $\bn$ for all signs of the cosmological constant.

The other set of solutions is given by the conditions $\alpha=0$, $\bn^2=-{1}/{\Lambda\beta^2}$ and the classical $r$-matrices \begin{align}
\label{rmatkappa2}
r=\frac{1}{\beta}(\Lambda J_a\otimes J^a+P_a\otimes P^a)+n^a\epsilon_{abc}(J^b\otimes P^c-P^c\otimes J^b)\qquad \bn^2=-\frac{1}{\Lambda\beta^2} \qquad \Lambda\neq 0.
\end{align}
It  exists only for $\Lambda\neq 0$ and corresponds to a Chern-Simons theory with gauge group $H_\Lambda$ and the non-gravitational $\Ad$-invariant symmetric bilinear form  \eqref{othform}. It  involves a vector $\bn$ which is timelike for $\Lambda<0$ (de Sitter case) and spacelike for $\Lambda>0$ (anti de Sitter case).  

This implies in particular that the usual $\kappa$-Poincar\'e symmetries given by the antisymmetric element \eqref{rakappa} with a {\em timelike} vector  $\bn=(\tfrac{1}{\kappa},0,0)$ are {\em not} compatible with the Chern-Simons formulation of (2+1)-gravity. The association  of timelike $\kappa$-Poincar\'e symmetries with (2+1)-gravity is possible only in the de Sitter case ($\Lambda<0$). The corresponding Chern-Simons theory is physically in-equivalent to (2+1)-gravity as it exhibits a different Poisson structure and becomes ill-defined in the limit $\Lambda\rightarrow 0$ in which its $\Ad$-invariant symmetric bilinear form \eqref{othform} degenerates.

\section{Outlook and Conclusions}
\label{conc}

In this article, we clarified the status of $\kappa$-Poincar\'e and Drinfeld doubles as quantum symmetries of  (2+1)-gravity and more general Chern-Simons theories based on the local isometry groups of (2+1)-gravity.  We showed how the question if a Hopf algebra arises as a quantum symmetry of the theory can be reduced to an explicit condition on the associated classical $r$-matrices and analysed this condition for the $\kappa$-Poincar\'e and Drinfeld doubles. 

As a result, we found that
the usual $\kappa$-Poincar\'e symmetries with a timelike vector as a deformation parameter are not compatible with (2+1)-gravity and can be associated with a Chern-Simons theory based on the isometry groups of (2+1)-gravity only in the de Sitter case ($\Lambda<0$). However, in this case, one does not obtain the Chern-Simons formulation of (2+1)-gravity but a physically in-equivalent Chern-Simons theory based on a different $\Ad$-invariant bilinear form and with different Poisson structure. This is in contrast to the situation for the Drinfeld doubles and {\em spacelike} $\kappa$-Poincar\'e symmetries, which are compatible with the Chern-Simons formulation of (2+1)-gravity for all signs of the cosmological constant.

In addition to clarifying the role of $\kappa$-Poincar\'e symmetries as symmetries of quantised (2+1)-gravity and Chern-Simons theory, these results suggest avenues for further research which could be pursued for either the timelike or spacelike version of these symmetries. In both cases, the association of the $\kappa$-Poincar\'e algebras with quantised Chern-Simons theory allows one to define models with $\kappa$-Poincar\'e symmetries on spacetimes with non-trivial topology.  Moreover, the possibility of  including point particles particles via minimal coupling to the Chern-Simons gauge field allows one to define non-trivial multi-particle systems  
with $\kappa$-Poincar\'e symmetries. This provides an explicit relation with an action functional as well as non-trivial particle interactions and therefore could prove useful for the physical interpretation of these symmetries as well as  the understanding of their role in quantum gravity.

\section*{Acknowledgements}

I thank Florian Girelli and Jerzy Jerzy Kowalski-Glikman for extensive discussions and correspondence.
This research was supported by Perimeter Institute for Theoretical
Physics.  Research at Perimeter Institute is supported by the Government
of Canada through Industry Canada and by the Province of Ontario through
the Ministry of Research \&
 Innovation.

\end{document}